\documentclass[graybox]{svmult}

\usepackage{helvet}         % selects Helvetica as sans-serif font
\usepackage{courier}        % selects Courier as typewriter font
\usepackage{type1cm}        % activate if the above 3 fonts are
\usepackage{makeidx}         % allows index generation
\usepackage{graphicx}        % standard LaTeX graphics tool
\usepackage{multicol}        % used for the two-column index
\usepackage[bottom]{footmisc}% places footnotes at page bottom
\usepackage{amsfonts}

\usepackage[pdf]{pstricks}

\usepackage{epstopdf}
\usepackage{graphicx, caption, subcaption}

\newcommand{\thf}[1]{\theta_p\left(#1\right)}
\newcommand{\pq}[2]{(pq)^{\frac{#1}{#2}}}

\begin{document}

\title*{Elliptic Integrable Models and Their Spectra from
Superconformal Indices }
\author{Anton Nedelin}
\institute{Anton Nedelin \at Section de Mathematiques, Universite de Geneve, 1211 Geneve 4, Switzerland and Department of Mathematics, King’s College London,
The Strand, WC2R 2LS London, United Kingdom \email{anton.nedelin@gmail.com}}

\maketitle

\abstract{In this contribution we summarize our recent progress in understanding the relation between ${\cal N}=1$ superconformal indices and 
relativistic elliptic integrable models. We start  briefly reviewing the emergence of such models in computations of the index in presence 
of surface defect. Next we give an example of such relation considering $4d$ theories obtained in the compactificaiton of $6d$ $(D_{N+3},D_{N+3})$ minimal 
conformal matter theories. In this case we obtain van Diejen model as well as its higher rank generalizations on $A_N$ and $C_2$ root systems. 
Finally we review a novel algorithm for computation of the ground states of elliptic integrable systems from superconformal indices that was recently proposed by us.}

\section{Introduction}
\label{sec:intro}

It is well known that supersymmetric gauge theories have an intricate relation with 
various integrable systems. This connection plays an important role in modern theoretical physics 
since it allows exact computations of certain observables in supersymmetric QFTs. Textbook example 
of this situation is the integrability of ${\cal N}=4$ Super Yang-Mills (SYM) theory
\cite{Minahan:2002ve,Beisert:2010jr}, which played an important role in solving the spectrum 
of this theory. On the other hand relations between QFTs and integrable systems can potentially 
shed light on some questions about the letter ones, and thus can be important for mathematics as well.

In the present contribution we will review our recent \cite{Nazzal:2021tiu,Nazzal:2023bzu,Nazzal:2023wtw}
progress in understanding one particular class of such relations namely the connection between $6d$ $(1,0)$ superconformal 
field theories (SCFTs) and elliptic relativistic quantum mechanics Hamiltonians in the spirit of Nekrasov and Shatashvili
\cite{Nekrasov:2009rc}. More specifically we will consider a class of $4d$ ${\cal N}=1$ theories obtained 
in the compactifications of various $6d$ SCFTs on a punctured Riemann surface. In order to obtain 
relation of these ${\cal N}=1$ theories with elliptic integrable systems we have to introduce surface defect into theory 
and compute the superconformal index of the resulting system.  

This construction was first proposed for the compactifications of $6d$ $(2,0)$ theory of ADE type \cite{Gaiotto:2012xa,Lemos:2012ph} 
where the corresponding integrable Hamiltonians appeared to be well known elliptic Ruijsenaars-Schneider (RS) analytic finite difference 
operators (A$\Delta$Os). Later these results were extended to several other examples \cite{Gaiotto:2015usa,Razamat:2018zel,Nazzal:2018brc}. 
In some of these cases resulting integrable systems were unknown to mathematicians. These results present high importance for physics 
since they can be used to compute superconformal indices of theories obtained in compactifications of the corresponding $6d$ theory on an arbitrary
Riemann surface \cite{Gadde:2009kb,Gadde:2011ik,Gadde:2011uv}. This computation technique is known as \textit{index bootstrap} and will be explained further in the text. 

In our recent works we have made several steps forward in the exploration of the connection between superconformal indices and 
elliptic integrable systems. In particular we have used recently obtained results on $4d$ compactifications of a family of $6d$ $(1,0)$ 
theories \cite{Kim:2018bpg,Razamat:2019ukg,Razamat:2020bix} as an input into construction of elliptic integrable systems. As a result 
we obtained van Diejen model as well as its higher rank generalizations which were previously not known. Additionally we have proposed 
an algorithm for the  derivation of the spectrum of general elliptic integrable system related to the index in the form of expansion in 
ellipticity parameters.

The contribution is organized as follows. In the next Section \ref{sec:elliptic:ops} we will briefly review the construction of elliptic integrable 
Hamiltonians from the superconformal indices. In the Section \ref{sec:van:Diejen} we  summarize the results obtained in \cite{Nazzal:2021tiu,Nazzal:2023bzu}. 
In particular we present novel higher rank generalizations of van Diejen integrable model that were derived in these papers.
Finally, in Section \ref{sec:spectrum} we give the main idea behind our novel algorithm for derivation of the spectrum of relativistic 
elliptic integrable operators related to superconformal indices which we proposed in \cite{Nazzal:2023wtw} and give an example
of application of this algorithm. 

\section{Elliptic A$\Delta$Os from superconformal indices.}
\label{sec:elliptic:ops}

We start with summarizing construction of relativistic elliptic integrable systems proposed in \cite{Gaiotto:2012xa}.
Starting point of this construction is some $6d$ $(1,0)$ SCFT $T_{6d}$  with the global symmetry $G_{6d}$. We put this 
theory on a Riemann surface ${\cal C}$ with some punctures, to be defined later, and turn on background gauge fields with fluxes 
${\cal F}$ preserving four supercharges. This triggers the RG flow accross dimensions down to $4d$, where we obtain some effective 
IR theory denoted $T_{4d}\left[ {\cal C},{\cal F} \right]$. It is often the case that $T_{4d}$ does not have Lagrangian description \cite{Gaiotto:2009we}. 

Important ingredients of our construction are punctures on Riemann surfaces. There exist various methods to define and classify punctures on compactification 
surfaces \cite{Gaiotto:2009we}. In our work we will limit ourselves with the class of $6d$ theories that have $5d$ effective gauge theory description 
with the gauge group $G_{5d}$. In this case we can see certain punctures, which we refer to as \textit{maximal punctures}, as superymmetry preserving boundary conditions 
imposed on fields of $5d$ effective theory. In particular Dirichlet boundary conditions imposed on $5d$ gauge fields  result in $4d$ gauge
theory acquiring extra factor $G_{5d}$ of the global symmetry which we associate to the puncture. At the same time $5d$ fields for which 
we impose Neumann boundary conditions give rise to $4d$ chiral operators $M$ naturally charged under $G_{5d}$. These special operators are referred to 
as \textit{moment maps}. In this approach punctures are characterized by the global symmetry $G_{5d}$ it brings to $4d$ theory and a natural set of 
moment map operators $M$. 
It is important to mention that for a given $T_{6d}$ effective $5d$ description is not necessarily unique since it is also 
defined by the holonomies of the $G_{6d}$ background field. Different choice of $5d$ description leads to different \textit{types} of 
maximal punctures one can have. 

There are two important operations one can perform on punctures of Riemann manifolds. First we can give vacuum expectation values (vevs) 
to the moment map operators of some puncture or their derivatives. Since moment map operators are charged under $G_{5d}$ it will obviously break $4d$ global 
symmetry associated with this puncture down to some subgroup. As a result we end up with the puncture corresponding to smaller global symmetry and less 
moment maps. This procedure is called \textit{partial closure of the puncture}. In case $G_{5d}$ is broken completely there is basically no puncture left 
after the procedure and we say that the \textit{puncture is closed}. In case the puncture can only be closed completely but not partially we call it 
\textit{minimal puncture}. Usually it comes with rank-one global symmetry, i.e. $\mathrm{U}(1)$ or $\mathrm{SU}(2)$. But examples with higher rank symmetries 
are also known \cite{Razamat:2018zel}. Importantly moment maps we give vev to can also be charged under $6d$ global symmetry 
$G_{6d}$ so that closure of puncture also results in the shift of the background fields flux ${\cal F}$ through the compactification surface.

Second operation we can perform with the puncture is \textit{surface gluing}. Assume we have two $4d$ theories: $T_{4d}^A$ obtained from some $T_{6d}$ 
in compactification on ${\cal C}_A$ surface with flux ${\cal F}_A$ and $T_{4d}^B$ obtained in compactification of the same $T_{6d}$ theory 
on another surface ${\cal C}_B$ with the flux ${\cal F}_B$ . The two surfaces can be completely different and have variety of punctures on them, but they both should have 
at least one maximal puncture of the same type. In this case we can glue two Riemann surfaces along this maximal puncture. At the level of 
$4d$ theories there are two possible procedures that realize this gluing. First one, called \textit{S-gluing}, corresponds to gauging the diagonal 
combination of the punctures global symmetries and identifying their moment map operators by adding superpotential $W\sim M\cdot M'$. In this case the 
flux through the resulting surface ${\cal C}_A\oplus{\cal C}_B$ is given by $\left({\cal F}_A-{\cal F}_B\right)$ (the overall sign of the flux is not 
a physical quantity). Second operation, referred to as \textit{$\Phi$-gluing}, also amounts to gauging the diagonal combination of $G_{5d}$. 
But identification of the moment maps this time requires an introduction of an extra chiral field $\Phi$ conjugate to one of $M$. For these 
fields we turn on superpotential $W\sim M\cdot\Phi-M'\cdot \Phi$. As the result of such gluing fluxes of two surfaces sum up resulting in 
$\left({\cal F}_A+{\cal F}_B \right)$. 

As we already explained in the introduction, the crucial field theory observable that we use to transform all the ideas explained above into 
the form of particular equations is the \textit{supersymmetric index} \cite{Kinney:2005ej,Romelsberger:2005eg,Rastelli:2016tbz} of a given $4d$ 
gauge theory. It is defined as the following trace over the Hilbert space of ${\cal N}=1$ theory quantized on $\mathbb{S}_3$
\begin{equation}
	{\cal I}(\{x_j\},q,p)= \mathrm{Tr}_{\mathbb{S}^3} (-1)^F \, q^{j_2-j_1+\frac{R}2} \, p^{j_2+j_1+\frac{R}2}\, \prod_{l=1}^{\mathrm{rank}\, G_F} x_l^{{\cal Q}_l}\,.
\end{equation}
Here $(j_1,j_2)$ are the two Cartan genrators of the $Spin(4)$ isometry of $\mathbb{S}_3$, $F$ is the fermion number, $R$ is ${\cal N}=1$ 
R-symmetry generator, ${\cal Q}_l$ are charges under the Cartan of the $4d$ global symmetry group $G_F$ of a given  ${\cal N}=1$ theory.
This includes both global symmetries associated to punctures and symmetries $G_{6d}$ inherited from the $6d$ theory. 
Variables $p$, $q$ and $x_l$ are called \textit{fugacities}. They help to keep track of charges of various BPS operators that index counts 
with respect to corresponding symmetries. Index is a function of all these fugacities and compactification geometry. 

Puncture operations explained above can be easily realized at the level of the index. Let's start from the gluing and 
assume once again that there are two $4d$ theories $T_{4d}^A$ and $T_{4d}^B$ obtained from the same $6d$ theory compactified on 
the Riemann surfaces ${\cal C}_A$ and ${\cal C}_B$ with fluxes ${\cal F}_A$ and ${\cal F}_B$ correspondingly.
Staring from the indices of these two theories we can construct the index corresponding to the compactification 
on the glued surface ${\cal C}_A\oplus{\cal C}_B$. For this purpose we  integrate, with certain measure, over fugacities of gauged global symmetry 
of the puncture we glue along:
\begin{eqnarray}
	{\cal I}\left[{\cal C}_A\oplus{\cal C}_B, {\cal F}_A\pm{\cal F}_B \right]\left(\{{\bf x}\}_{A\cup B }\right)=\oint \prod\limits_{i=1}^{\mathrm{rank} G_{5d}}
\frac{dz_i}{2\pi iz_i}\Delta_{\mathrm{Haar}}\left(\{z\}\right)\times\nonumber\\
\Delta^{S/\Phi}\left(\{z\} \right){\cal I}\left[ {\cal C}_A,{\cal F}_A \right]\left(\{x\}_A\cup \{z\}\right){\cal I}
\left[ {\cal C}_B,{\cal F}_B \right]\left(\{x\}_B\cup \{z\}\right)\,.
\label{index:gluing}
\end{eqnarray}
Here $\Delta_{\mathrm{Haar}}$ is the Haar measure of the puncture symmetry $G_{5d}$, $\Delta^{S/\Phi}$ is the contribution of the gauge fields 
of $G_{5d}$ and, in case of $\Phi$-gluing, chiral multiplet $\Phi$. As previously explained fluxes are summed in case of S-gluing and 
subtracted in case of $\Phi$-gluing. 

Finally we should also explain what does closing a puncture make to the index. As explained previously in order to close the puncture 
we give vev to one of the moment map operators or its derivatives. To understand an effect of this on the index let's consider 
simplified situation of a theory with some $\mathrm{U}(1)_u$ global symmetry, $u$ being corresponding fugacity.
Assume we give vev to some operator ${\cal O}$ that has charge $-1$ under this symmetry and contributes to the index with the weight
$U^*\cdot u^{-1}=q^{j_2^{ {\cal O} }-j_1^{ {\cal O} }+\frac{1}{2}R^{\cal O}}p^{j_2^{ {\cal O} }+j_1^{ {\cal O} }+\frac{1}{2}R^{\cal O}}u^{-1}$. 
This vev triggers an RG flow which arrives to the IR theory with the index given by \cite{Gaiotto:2012xa} 
\begin{equation}
	{\cal I}_{IR}\sim\mathrm{lim}_{u\to U^*}{\cal I}_{UV}(u)\,,
\label{index:vev}
\end{equation}
where we omit some overall factors that will not play role in our calculations. 

Now we have in our hands all ingredients to explain the emergence of integrable systems in index calculations. 
Starting point of the calculation are two theories. First one is  obtained from certain $T_{6d}$ in compactifications on a general 
Riemann surface ${\cal C}_{g,s}[u]$ of genus $g$ with $s$ punctures at least one of which is  maximal puncture parametrized by fugacity $u$
and flux ${\cal F}$.
Second theory is obtained from  the same $T_{6d}$ compactified on a sphere with three punctures, one minimal and two maximal 
of the same type as $u$-puncture of ${\cal C}_{g,s}$, and some flux ${\cal A}$. This surface plays an important role in the studies of $6d$ compactifications and 
is called \textit{trinion}. We denote it with ${\cal T}_{u,v,\hat{z}}^{{\cal A} }$, where $u$ and $v$ are fugacities of the maximal punctures and 
$\hat{z}$ is that of the minimal puncture. An important assumption we have to make is that $4d$ theory obtained in such compactification 
is known and has Lagrangian description, which will always be the case for examples we work with.

Now as a first step of our construction we glue trinion ${\cal T}_{u,v,\hat{z}}$ and Riemann surface ${\cal C}_{g,s}[u]$ together 
along $u$-puncture and compute the index ${\cal I}\left[{\cal C}_{g,s}[u]\oplus{\cal T}^{ {\cal A}}_{u,v,\hat{z}},{\cal F}+{\cal A}\right]$ 
obtained in the compactification on such a composite surface. Next we close the minimal ${\hat z}$ puncture by giving vev to 
some the holomorphic derivatives\footnote{We introduce two complex coordinates $z_1=x_1+ix_2$ and 
$z_2=x_3+ix_4$ on $\mathbb{R}_4$.} $\partial_1^{r}\partial_2^{m}{\hat M}$ of one of its moment map operators ${\hat M}$. Assuming 
weight of the operator ${\hat M}$ in the index is $U^*$ in the IR one usually obtains the following identity
\begin{eqnarray}
	{\cal I}_{IR}^{(\hat{z},\hat{M};r,m)}\left[{\cal C}_{g,s}[v],{\cal F} \right]\sim
	\mathrm{lim}_{{\hat z}\to p^rq^mU^*}{\cal I}\left[ {\cal C}_{g,s}[u]\oplus{\cal T}^{ {\cal A}}_{u,v,\hat{z}},{\cal F}+{\cal A} \right]\sim 
	\nonumber\\
	{\cal O}_u^{(\hat{z},\hat{M};r,m) }\cdot{\cal I}\left[{\cal C}_{g,s}[u],{\cal F} \right]\,,
		\label{ado:def}
\end{eqnarray}
where ${\cal O}_u^{(\hat{z}, \hat{M};r,m) }$ is a tower of relativistic elliptic Hamiltonians. They all act on the $G_{5d}$ fugacity $u$ and 
labeled by the type of minimal puncture we close ($\hat{z}$), moment map operator $\hat{M}$ we use for it  as well as integers $r$ and $m$ corresponding to 
the powers of the holomorphic derivatives in the vev. Left hand side of the equation contains index of an IR theory which we also label 
by the same set of parameters since theory we arrive to in IR depends on all this information. Second expression just corresponds to gluing and 
closing puncture described above and final expression on the right hand side is the result of the calculation. Just as before similarity signs mean 
that we omit some of the overall constants which are irrelevant for our purposes. It is also important to mention that as can be seen from Eq.(\ref{ado:def}) 
geometrically after all operations we arrive to the original Riemann surface ${\cal C}_{g,s}$ with the same flux ${\cal F}$ but with the defect introduced. 
In order for this to work the shift in flux resulting from the closure of $\hat{z}$ puncture should exactly compensate trinion flux ${\cal A}$. 
Trinion with such flux can always be constructed by appropriate manipulations\cite{Nazzal:2021tiu}. Finally we should clarify that 
result in Eq. (\ref{ado:def}) is an expectation based on the computations performed in many different examples and it does not have 
any proof. The same result in some cases was reproduced by the direct computation of the index in presence of a codimension two defect\cite{Gadde:2013dda}.

Before moving towards implementations of the algorithm summarized above we should mention two crucial properties 
of the obtained ${\cal O}$ operators. Both of these properties are consequences of the S-duality between $4d$ theories. 
They can also be seen from some geometric arguments applied to the compactificatiton surface. For example starting with the 
surface having two minimal punctures and closing them one by one we obtain product of two operators similar to one in Eq.(\ref{ado:def}). 
But the order of these two operators in the product depends on the order we close punctures in. However index should be insensitive to 
this order and hence we can conclude  that all such operators should commute with each other:
\begin{equation}
	\left[{\cal O}_u^{(\hat{a},\hat{M};r,m)},{\cal O}_u^{(\hat{a}',\hat{M}';r',m')} \right]=0\,.
\end{equation} 
This signals about integrability of the obtained system, since at once we obtain a full tower of the commuting 
operators.

Second property can also be deduced from the S-duality or similar geometric arguments. The claim is that index of 
any $4d$ theory emerging  in the compactification of a given $T_{6d}$ is the \textit{Kernel function} for the 
observed operators ${\cal O}$. In particular the following holds: 
\begin{equation}
	{\cal O}_u^{(\hat{a},\hat{M};r,m)}\cdot {\cal I}\left[{\cal C}_{g,s}[u,v],{\cal F} \right]=
	{\cal O}_v^{(\hat{a},\hat{M};r,m)}\cdot {\cal I}\left[{\cal C}_{g,s}[u,v],{\cal F} \right]\,.
	\label{kernel:property}
\end{equation}
In this equation operators on two sides act on the fugacities of different punctures. These two punctures can be of the same type 
in which case the operators are the same or of different types in which case operators are also different. This Kernel function property 
of operators and indices plays crucial role in the calculations of latter ones \cite{Gaiotto:2012xa, Gadde:2011uv}. In Section \ref{sec:spectrum} 
we will return to this question and propose a construction of the spectra of elliptic operators relying on this property.

\section{Results: van Diejen model and its generalizations}
\label{sec:van:Diejen}

In the present section we will briefly summarize some particular results of applying presented 
construction to the compactifications of some $6d$ SCFTs. The $6d$ theory that was the starting point of 
our investigation is so-called \textit{minimal $(D_{N+3},D_{N+3})$ conformal matter theory} which is a SCFT with $(1,0)$ 
supersymmetry. Physically this theory arises as a world-volume theory of an $M5$ brane probing $D_{N+3}$ singularity. 
It has $G_{6d}=\mathrm{SO}(4N+12)$ global symmetry which enhances to $E_8$ symmetry in $N=1$ case. In this latter case 
one obtains $6d$ SCFT called \textit{E-string theory}. As explained in the previous Section \ref{sec:elliptic:ops} we define 
type of punctures we can put on a compactification surface through the effective $5d$ gauge theory description of the $6d$ SCFT. 
In case of the minimal conformal matter there are three possible $5d$ theories with $\mathrm{SU}(N+1)$, $\mathrm{USp}(2N)$ and $\mathrm{SU}(2)^N$ gauge 
groups. These leads to three possible types of maximal punctures. In our works so far we have only considered $\mathrm{SU}(N+1)$ and $\mathrm{USp}(2N)$ punctures leaving 
$\mathrm{SU}(2)^N$ puncture for the future research due to technical difficulties. Minimal puncture in both cases we consider is 
just an $\mathrm{SU}(2)$ puncture obtained by partially closing $\mathrm{USp}(2N)$ maximal one. 

\begin{figure}[b]
	\begin{subfigure}[b]{0.45\textwidth}
		\includegraphics[width=\textwidth]{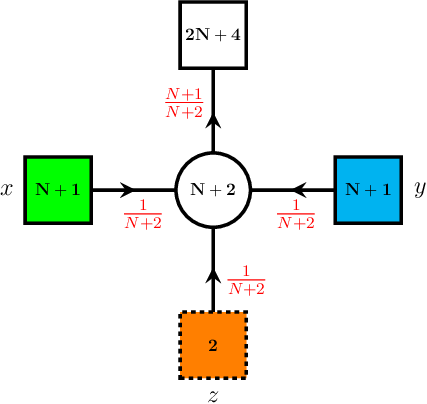}
	\caption{Trinion compactification with two $\mathrm{SU}(N+1)$ maximal and one $\mathrm{SU}(2)$ minimal punctures.}
	\end{subfigure}
	\hspace{7mm}
	\begin{subfigure}[b]{0.45\textwidth}
	\includegraphics[width=\textwidth]{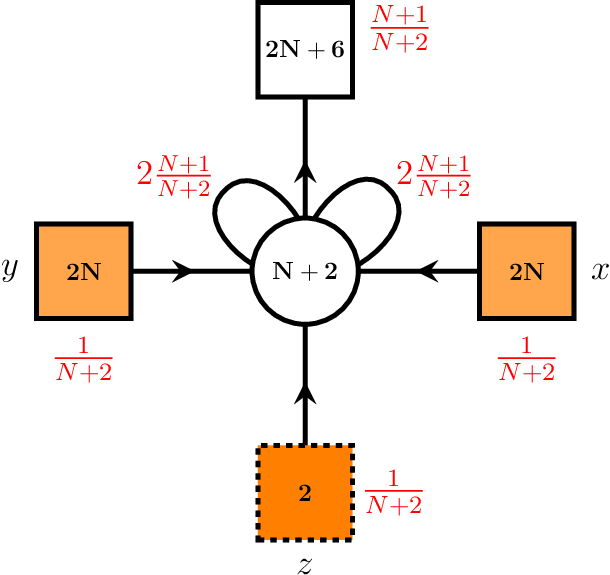}
	\caption{Trinion compactification with two $\mathrm{USp}(2N)$ maximal and one $\mathrm{SU}(2)$ minimal punctures.}
\end{subfigure}
\caption{Quivers of trinion compactifications with $\mathrm{SU}(N+1)$ and $\mathrm{USp}(2N)$ maximal punctures. As usually circles 
stand for ${\cal N}=1$ gauge multiplets, squares for matter multiplets in fundamental/antifundamental representations and 
loops  for matter multiplets in rank-two antisymmetric representation. We also specify $R$-charges of multiplets in red.}
\label{pic:trinions}
\end{figure}

As mentioned previously crucial element we need for 
our construction is the trinion compactification with known Lagrnagian description. Trinion compactifications we used in our projects were derived in 
\cite{Razamat:2020bix} and \cite{Nazzal:2021tiu} and are given by relatively simple $4d$ gauge theories. In particular trinion compactification 
with two $\mathrm{SU}(N+1)$ maximal and one $\mathrm{SU}(2)$ minimal punctures is given by $\mathrm{SU}(N+2)$ ${\cal N}=1$ SQCD with $(2N+4)$ flavors. 
Trinion with $\mathrm{USp}(2N)$ maximal punctures is given by $\mathrm{SU}(N+2)$ ${\cal N}=1$  gauge theory with $(4N+2)$ fundamental, $(2N+6)$ antifundamental 
and $2$ rank-two antisymmetric multiplets. Quivers of these theories are shown on Figure \ref{pic:trinions}.

Starting from these theories we can perform all the calculations for the derivation of the corresponding finite difference operators. 
Details of these calculations can be found in our recent papers \cite{Nazzal:2021tiu} for $\mathrm{SU}(N+1)$ punctures and 
\cite{Nazzal:2023bzu} for $\mathrm{USp}(2N)$ punctures. Here we just briefly summarize most important of the obtained results. 

In case of $\mathrm{SU}(N+1)$ type punctures, which we also refer to as $A_N$ type, we have obtained the tower of finite difference 
operators. The action of the \textit{basic operator} of the tower corresponding to the choice $r=0$ and $m=1$ in Eq.(\ref{ado:def}) is given by 
\begin{eqnarray}
	&&{\cal O}_{x}^{(A_N;h_i^{-1};1,0)} \cdot {\cal I}(x)\equiv
\nonumber\\
&&\hspace{.1cm}\left(\sum\limits_{l\neq m}^{N+1}A^{(A_N;h_i^{-1};1,0)}_{lm}(x)\Delta_q^{-1}(x_l)\Delta_q(x_m)+ W^{(A_N;h_i^{-1};1,0)}\left(x,h\right)\right) {\cal I}(x)\,,
\label{sun:op:noflip:gen}
\end{eqnarray}
where $\Delta_q(x)$ is operator performing the shift $x\to q x$,  $A_{lm}$ are  functions given by the following expression:
\begin{eqnarray}
&&A^{(A_N;h_i^{-1};1,0)}_{lm}(x)=\frac{\prod\limits_{j=1}^{2N+6}\thf{\pq{1}{2} h_j^{-1}x_l^{-1}}}
{\thf{\frac{x_m}{x_l}}\thf{q\frac{x_m}{x_l}}}\times\nonumber\\
&&\hspace{1.3cm}\prod\limits_{k\neq m\neq l}^{N+1}\frac{\thf{\pq{1}{2}h_i^{-1}x_k^{-1}}
\thf{\pq{1}{2}h_i^{-1}h^{1/2} x_k}}{\thf{\frac{x_k}{x_l}}\thf{\frac{x_m}{x_k}}}\,,
\end{eqnarray}
and $W^{(A_N;h_i^{-1};1,0)}\left(x,h\right)$ is quite complicated function elliptic in all variables $x$. Details of this function 
as well as the full tower of operators can be found in \cite{Nazzal:2023bzu}. Finally $\thf{x}$ is the standard $p$-theta function. 
Operators act on the $N$ fugacities $x_i$ of the $\mathrm{SU}(N+1)_x$ puncture. 
They depend on $(2N+6)$ parameters $h_i$ with $h=\prod_{i=1}^{2N+6}h_i$. Physically these parameters correspond to the charges of the moment maps of the 
maximal puncture. In total there are $(2N+6)$ operators parametrized by the choice of the moment map of the minimal $\mathrm{SU}(2)$ puncture we give vev to. 

In the special case of $N=1$ when original $6d$ SCFT becomes E-string theory all operators (\ref{sun:op:noflip:gen}) reduce to the 
\textit{van Diejen} model \cite{van:diejen}
\begin{equation}
	{\cal O}_x\cdot{\cal I}(x)=\left[\frac{\prod\limits_{i=1}^8\thf{\pq{1}{2}h_i^{-1}x}}{\thf{qx^2}\thf{x^2}}{\cal I}(qx)+(x\to x^{-1})\right]+W(x;h_i){\cal I}(x)\,,
	\label{vd:model}
\end{equation}
where the second term is the same as the first one but with the substitution of $x\to x^{-1}$. Our result also appears to be consistent with the 
previous observation of relation between van Diejen model and compactifications of E-string theory \cite{Nazzal:2018brc} which used different trinion 
theory as a starting point. This model is a well studied example of relativistic elliptic integrable systems and was known to mathematicians . 
Our operator presented in (\ref{sun:op:noflip:gen}) in turn is a higher rank generalization of van Diejen model defined on $A_N$ root system. 
Unlike van Diejen model, this generalization was not previously known.  

In addition we have studied compactifications on the surfaces with $\mathrm{USp}(2N)$ (type $C_N$)  maximal punctures using the trinion theory 
shown on the Figure \ref{pic:trinions}. In this case we unfortunately stumbled upon technical difficulties and were only able to perform 
calculations in cases of $C_1$ and $C_2$ punctures. In the former case we one more time landed on the van Diejen model (\ref{vd:model}) 
described in a slightly different yet equivalent way. In $C_2$ case on the other hand we managed to find corresponding higher rank generalization 
of van Diejen Hamiltonian. The lowest operator of the tower corresponding to the choice $r=0$ and $m=1$ in Eq.(\ref{ado:def}) is given by 
\begin{eqnarray}
	&&{\cal O}^{(C_2;h_k;1,0)}_x =\sum\limits_{i=1}^2\left(A^{(C_2;h_k;1,0)}_i(x,h)\Delta_q(x_i)+A^{(C_2;h_k;1,0)}_i(x^{-1},h)\Delta_q(x_i)^{-1}\right)
	\nonumber\\
	&&\hspace{7cm}+W^{(C_2;h_k;1,0) }(x,h)\,,
\label{c2:operator}
\end{eqnarray}
where $A_i$ functions are given by 
\begin{eqnarray}
&&A^{(C_2;h_k;1,0)}_i(x,h)=\prod\limits_{j\neq i}^2\frac{1}{\thf{x_i^2}\thf{qx_i^{2}}\thf{x_ix_j^{\pm 1}}}\times
\nonumber\\
&&\hspace{3cm}\thf{\pq{1}{2}h_kx_j^{\pm 1}}
\prod\limits_{l=1}^{10}\thf{\pq{1}{2}h_l^{-1}x_i}\,,
\end{eqnarray}
and $W^{(C_2;h_k;1,0) }(x,h)$ is the function elliptic in all parameters with the poles located at $x_i=sq^{\pm 1/2}$, $x_i=sp^{1/2}q^{\pm 1/2}$ where 
$s=\pm 1$. Just as in $A_N$ case operators in Eq.(\ref{c2:operator}) depend on the charges $h_i$ of the moment map operators of the maximal puncture. 
In this case $\mathrm{USp}(4)$ puncture has $10$ moment maps operators. Accordingly there are $10$ ways to close the minimal puncture as well and 
hence we have $10$ operators in total which we label by $h_k$. 

In addition to deriving generalizations (\ref{sun:op:noflip:gen}) and (\ref{c2:operator}) of the van Diejen model, we have also checked some of 
their basic properties. In particular we have checked commutation relations of these operators using expansion in $p$ and $q$. Importantly 
we also checked the kernel property of some  ${\cal N}=1$ indices with respect to the corresponding operators. In particular one of the 
most interesting results presented in \cite{Nazzal:2023bzu} is an analytic proof of the kernel property for the function of $A_2C_2$ type. For this we considered 
compactification on a sphere with two maximal punctures, one with an $\mathrm{SU}(3)$ and one with $\mathrm{USp}(4)$ symmetry. According to our 
proposition the index of resulting $4d$ theory should satisfy kernel property as in Eq.(\ref{kernel:property}) with $A_2$ operator 
(\ref{sun:op:noflip:gen}) on one side and $C_2$ operator (\ref{c2:operator}) on the other side of equation. In our paper we check that 
this proposition indeed holds by the direct computation of the two sides.

\section{Spectra of elliptic integrable systems from large compactifications.}
\label{sec:spectrum}

In the previous sections we summarized derivation of the elliptic integrable systems from the superconformal indices and 
gave examples of novel operators emerging in this construction. In the present section we will summarize a novel algorithm 
of the derivation of at least ground states of these type of operators using their relation to indices of supersymmetric gauge theories. 
This algorithm was first proposed by  us in \cite{Nazzal:2023wtw} and more detailed explanation of it can be found there. 

Crucial role in our argument is played by the kernel property (\ref{kernel:property}) of $4d$ ${\cal N}=1$ indices. Let's say we want to 
find the spectrum of our elliptic operators
\begin{equation}
	{\cal O}_{ {\bf x}}^\alpha\cdot\psi_{\lambda_i}( {\bf x})=E_{\alpha,\lambda_i}\psi_{\lambda_i}( {\bf x})\,,
	\label{spectrum:def}
\end{equation}
where we have encoded all possible labels of operators in the tower in $\alpha$ index and labeled eigenvalues by 
$\lambda_i$ which is usually a partition rather then just an integer.
We also assume that there is a natural ordering of these spectrum labels so we can enumerate them. Label $\lambda_0$ in this case corresponds 
to the ground state. 

An important fact that directly follows from the 
kernel property is that the index of the theory obtained in the compactifcation of some $T_{6d}$ on the Riemann surface ${\cal C}$ with 
$s$ maximal punctures of the same type can be written as the following diagonal expansion in the basis of $\psi_\lambda$ 
\begin{equation}
	{\cal I}\left[T_{6d},{\cal C} \right](\{ {\bf x_j} \} )=\sum\limits_{i=0}^{\infty}C_{\lambda_i}\prod\limits_{j=1}^s\psi_{\lambda_i}({\bf x_j})\,,
	\label{expansion}
\end{equation}
where coefficients $C_{\lambda_i}$ can depend on the fugacities of $6d$ global symmetry $G_{6d}$ as well as $p$ and $q$. This expression 
relies on the kernel property of indices and follows from the construction known as \textit{index bootstrap} \cite{Gaiotto:2012xa}. It plays 
an important role in running index computations, especially in the cases when $4d$ theories do not have Lagrangian descriptions and 
other computational methods fail. 

Now let's assume we know the  compactification of the theory on some surface ${\cal C}$ with at least two maximal punctures and the  flux ${\cal F}$. Corresponding 
index then can be written as follows
\begin{equation}
{\cal I}_1\left({\bf x}_1, {\bf x}_2 \right)=\sum\limits_{i=0}^{\infty}C_{\lambda_i}\psi_{\lambda_i}({\bf x}_1)\psi_{\lambda_i}({\bf x}_2)
	\label{tube}
\end{equation}
In principle ${\cal C}$ can have other punctures and some genus in which case all this information is hidden in the coefficients $C_{\lambda_i}$ of the expansion. 
Now let's sequentially glue $n$ copies of ${\cal C}$ along maximal punctures as specified in Eq.(\ref{index:gluing}). 
Resulting surface will have once again two maximal punctures, flux $n{\cal F}$ and number of other punctures as well
as genus also multiplied by $n$. The index  of the resulting compactification theory is given by
\begin{equation}
	{\cal I}_n\left({\bf x}_1, {\bf x}_2 \right)=\sum\limits_{i=0}^{\infty}\left(C_{\lambda_i}\right)^n\psi_{\lambda_i}({\bf x}_1)\psi_{\lambda_i}({\bf x}_2)
	\label{n:tube}
\end{equation}
Our goal now is to find the ground state wavefunction $\psi_{\lambda_0}$ and corresponding energy $E_{\lambda_0}$ in expansion in $p$ and $q$ fugacities. 
This expansion should make sense since index itself is always regular in such an expansion. We also assume that the lowest order of expansion of $C_{\lambda_i}$ 
coefficients grows with $i$. This assumption is supported by empiric observation made in \cite{Gaiotto:2012xa} in case of the compactification of $A_1$ $(2,0)$ 
theory where the corresponding elliptic operator is $A_1$ Ruijsenaar-Schneider model and the spectrum is known in certain limits. 
Also we check our conjecture for consistency in the end of calculations. Under this assumption we can take the limit $n\to\infty$ and conclude that in this limit 
\textit{only the ground state contributes} and we can immediately write down $C_{\lambda_0}$ coefficient
\begin{equation}
\label{Eq:C0}
C_{0}\equiv C_{\lambda_0}= \lim_{n\to \infty} \frac{{\cal I}_{n+1} ({\bf x}_1,{\bf x}_2)}{{\cal I}_{n} ({\bf x}_1,{\bf x}_2)}\,.
\end{equation}
From here we can also obtain the corresponding ground state eigenfunction
\begin{equation}
\label{Eq:psi0}
\widetilde \psi_0({\bf x})\equiv \psi_0({\bf 1})\; \psi_0({\bf x})= \lim_{n\to \infty}\frac{1}{\left(C_0\right)^{n}}\,
{\cal I}_{n} ({\bf x},1)\,.
\end{equation}
Both equations above are in principle exact. In practice however taking $n\to\infty$ limit is an extremely difficult problem. So in our calculations 
we derive both quantities as expansion in $p$ and $q$ fugacities. In this case since lowest order of this expansion of $C_{\lambda_i}$ and hence 
also of $\psi_{\lambda_i}$ grows with $i$ we can conclude that to a given order of expansion equations (\ref{Eq:C0}) and (\ref{Eq:psi0}) work for 
some large but finite $n$. In this way we can perform particular calculations. Repeating similar argument we can also derive spectrum of 
elliptic operators beyond the ground state \cite{Nazzal:2023wtw}. 

In \cite{Nazzal:2023wtw} we have tested this algorithm for many models. One particular example we went through is van Diejen model (\ref{vd:model}) 
with all parameters $h_i$ taken to be equal, i.e. $h_i=t~\forall ~i=1,\cdots,8$. The ground state eigenfunction appeared to be given by 
\begin{eqnarray}
\widetilde \psi_0(x) = 1+\sqrt{pq}\left[16 t^{-1}+8t \left(x+\frac{1}{x}\right)\right]
+pq\left[108 t^{-2}-69
\right. \nonumber\\\left.
+128 \left(x+\frac{1}{x}\right)-
\left(x^2+\frac{1}{x^2}\right)+36t^2\left(1+x^2+\frac{1}{x^2}\right) \right]
\nonumber\\
+\sqrt{pq}(p+q)\left[16 t^{-1}+8t \left(x+\frac{1}{x}\right) \right]+\cdots\,.
\label{vd:ground:state}
\end{eqnarray}
Directly substituting this eigenfunction into the eigenvalue equation we can also extract ground state energy to the 
first orders in $p$ and $q$ expansion
\begin{equation}
	E_0=1-p-q+\cdots
\label{vd:ground:state:energy}
\end{equation}
In \cite{Nazzal:2023wtw} we have also derived ground state and first excited state of $A_1$ Ruijsenaars-Schneider model as well 
as Razamat models proposed in \cite{Razamat:2018zel}. In all of the cases algorithm has proven to work correctly. 

\section{Discussion and Outlook.}
\label{sec:discussion}

In this contribution we gave a brief review of the recent progress \cite{Nazzal:2021tiu,Nazzal:2023bzu,Nazzal:2023wtw} 
in understanding relativistic elliptic integrable systems from 
the point of view of $4d$ superconformal indices. In particular we have reviewed construction of such operators from the 
theories obtained in compactifications of $6d$ SCFTs on punctured Riemann surfaces. We also presented two classes of novel operators 
of this kind. And finally we presented a novel algorithm for the derivation of the spectra (or at least ground state) of these operators 
as an expansion in ellipticity parameters. 

Our findings open up a promising field of research with many potential directions to investigate. First of all 
we can extend our construction to other examples of compactifications of various $(1,0)$ theories. This includes 
$C_N$ and $(A_1)^N$ punctures of the minimal $(D,D)$ conformal matter theory, rank Q E-string theory \cite{Hwang:2021xyw,Pasquetti:2019hxf} 
and non-minimal conformal matter theory \cite{Heckman:2015bfa}. Second important direction is investigation of spectra of known and 
new operators. So far we have tested our novel algorithm on a limited set of models (mostly low rank) and for the ground states only. Improving 
our code we can break through both of these limitations. Finally using the information about the spectrum we can come back to the computations of the indices 
and implement an effective method of evaluations for ${\cal N}=1$ indices even when the latter one lack Lagrangian description in a same way it was 
previously done for ${\cal N}=2$ indices.

%\begin{figure}[b]
%\sidecaption
% For example, with the graphicx style use
%\includegraphics[scale=.65]{figure}
%
% If no graphics program available, insert a blank space i.e. use
%\picplace{5cm}{2cm} % Give the correct figure height and width in cm
%
%\caption{If the width of the figure is less than 7.8 cm use the
%\texttt{sidecapion} command to flush the caption on the left side of
%the page. If the figure is positioned at the top of the page, align
%the sidecaption with the top of the figure -- to achieve this you
%simply need to use the optional argument \texttt{[t]} with the
%\texttt{sidecaption} command}
%\end{figure}

%\begin{figure}[t]
%\sidecaption[t]
%% Use the relevant command for your figure-insertion program
%% to insert the figure file.
%% For example, with the option graphics use
%\includegraphics[scale=.65]{figure}
%%
%% If no graphics program available, insert a blank space i.e. use
%%\picplace{5cm}{2cm} % Give the correct figure height and width in cm
%%
%%\caption{Please write your figure caption here}
%\caption{If the width of the figure is less than 7.8 cm use the
%\texttt{sidecapion} command to flush the caption on the left side of
%the page. If the figure is positioned at the top of the page, align
%the sidecaption with the top of the figure -- to achieve this you
%simply need to use the optional argument \texttt{[t]} with the
%\texttt{sidecaption} command}
%\end{figure}

%
\begin{acknowledgement}
I would like first of all to acknowledge my collaborators Shlomo Razamat and Belal Nazzal together with whom I accomplished the series of works this contribution is based on. 
I am also grateful to Luca Cassia and Alba Grassi with whom I discussed these projects while working on them. The research presented in this contribution was supported 
by the Swiss National Science Foundation, Grant No. 185723 and STFC grant number ST/X000753/1. 
\end{acknowledgement}

\end{document}